\documentclass[a4paper,english]{article}
\usepackage[T1]{fontenc}
\usepackage[latin1]{inputenc}
\usepackage{graphicx}

\makeatletter

\usepackage{graphicx}

\usepackage{babel}
\makeatother
\date{}
\begin{document}

\title{The conversion of Neutron star to Strange star : A two step process}

\maketitle
\begin{center}{\large Abhijit Bhattacharyya$^{\textrm{1}}$, Sanjay
K. Ghosh$^{\textrm{2,3}}$, Partha S. Joardar$^{\textrm{3}}$, Ritam Mallick$^{\textrm{2}}$and
Sibaji Raha$^{\textrm{2,3}}$}\end{center}{\large \par}

{\large \par{}}{\large \par}

\begin{center}$^{\textrm{1}}$Department of Physics, University of
Calcutta, 92, A.P.C Road, Kolkata - 700009, INDIA\end{center}

\begin{center}$^{\textrm{2}}$Department of Physics, Bose Institute,
93/1, A.P.C Road, Kolkata - 700009, INDIA\end{center}

\begin{center}$^{\textrm{3}}$Centre for Astroparticle Physics and
Space Science, Bose Institute, 93/1, A.P.C Road, Kolkata - 700009,
INDIA\end{center}

\begin{abstract}
The conversion of neutron matter to strange matter in a neutron star
have been studied as a two step process. In the first step, the nuclear
matter gets converted to two flavour quark matter. The conversion
of two flavour to three flavour strange matter takes place in the
second step. The first process is analysed with the help of equations
of state and hydrodynamical equations, whereas, in the second process,
non-leptonic weak interaction plays the main role. Velocities and
the time of travel through the star of these two conversion fronts
have been analysed and compared. 
\end{abstract}
%\maketitle

\section*{I. Introduction }

It has been conjectured that strange quark matter, consisting of almost
equal numbers of u, d and s quarks, may be the true ground state of
strongly interacting matter \cite{key-1,key-2} at high density and/or
temperature. This conjecture is supported by bag model calculations
\cite{key-3} for certain range of values for the strange quark mass
and the strong coupling constant. By considering realistic values
for the strange quark mass (150 - 200 MeV \cite{key-4}), it may be
shown that the strangeness fraction in a chemically equilibrated quark
matter is close to unity for large baryon densities. Such bulk quark
matter would be referred to as \char`\"{}strange quark matter (SQM)\char`\"{} 
in what follows.

The above hypothesis may lead to important consequences both for laboratory
experiments as well as for astrophysical observations. Normal nuclear
matter at high enough density and/or temperature, would be unstable
against conversion to two flavour quark matter. The two flavour quark
matter would be metastable and would eventually decay to SQM, releasing
a finite amount of energy in the process. Such conversion may take
place in the interior of a neutron star where the densities can be
as high as (8-10)$\rho_{0}$ with $\rho_{0}$ being the nuclear matter
density at saturation \cite{key-5,key-6}. If Witten's conjecture
\cite{key-1} is correct, the whole neutron star may convert to a
strange star with a significant fraction of strange quarks in it.
(Neutron star may also become a hybrid star with a core of SQM in
case the entire star is not converted to a strange star. Such a hybrid
star would have a mixed phase region consisting of both the quark
matter and the hadronic matter \cite{key-7}.) Hadron to quark phase
transition inside a compact star may also yield observable signatures
in the form of Quasi-Periodic Oscillations (QPO) and the Gamma ray
bursts \cite{key-7a,key-7b}.

There are several ways in which the conversion may be triggered at
the centre of the star. A few possible mechanisms for the production
of SQM in a neutron star have been discussed by Alcock {\it et al} \cite{key-8}.
The conversion from hadron matter to quark matter is expected to start
as the star comes in contact with a seed of external strange quark
nugget. Such a seed would then grow by 'eating up' baryons in the
hadronic matter during its travel to the centre of the star, thus
converting the neutron star either to a strange star or a hybrid star.
Another mechanism for the initiation of the conversion process was
given by Glendenning \cite{key-7} . It was suggested there that a
sudden spin down of the star may increase the density at its core,
thereby triggering the conversion process spontaneously.

Conversion of neutron matter to strange matter has been studied by
several authors. Olinto \cite{key-9} viewed the conversion process
to proceed via weak interactions as a propagating slow-combustion
(i.e. a deflagration) front and derived the velocity of such a front.
Olsen and Madsen \cite{key-10} and Heiselberg {\it et al} \cite{key-11}
estimated the speed of such conversion front to range between 10 m/s
to 100 km/s. The combustive conversion front was assumed to have a
microscopic width of a few tens to a few hundreds of fm in these calculations.

Collins and Perry \cite{key-12}, on the other hand, assumed that
the hadronic matter gets converted first to a two flavour quark matter
that eventually decays to a three flavour strange matter through weak
interactions. Lugones {\it et al} \cite{key-13} argued that the hadron
to SQM conversion process may rather proceed as a detonation than
as a deflagration even in the case of strangeness production occuring
through seeding mechanisms \cite{key-8}.

Horvarth and Benvenuto \cite{key-14} examined the hydrodynamic stability
of the combustive conversion in a non-relativistic framework. These
authors inferred that a convective instability may increase the velocity
of the deflagration front, so that a transition from slow combustion
to detonation may occur. They argued that such a detonation may as
well be responsible for the type II supernova explosions\cite{key-15}.
In a relativistic framework, Cho {\it et al} \cite{key-16} examined the
conservation of the energy-momentum and the baryonic density flux
across the conversion front. Using Bethe-Johnson \cite{key-16a} and
Fermi-Dirac neutron gas \cite{key-16b} equations of state (EOS) for
the nuclear matter (NM) and the Bag model for SQM, they found that
the conversion process was never a detonation but a slow combustion
only for some special cases. Recently, Tokareva {\it et al} \cite{key-17}
modelled the hadron to SQM conversion process as a single step process.
They argued that the mode of conversion would vary with temperature
of SQM and with the value of bag constant in the Bag model EOS. Berezhiani
{\it et al} \cite{key-18}, Bombaci {\it et al} \cite{key-19} and Drago {\it et al}
\cite{key-20}, on the other hand, suggest that the formation of SQM
may be delayed if the deconfinement process takes place through a
first order transition \cite{key-20a} so that the purely hadronic
star can spend some time as a metastable object.

In this paper, we model the conversion of nuclear matter to SQM in
a neutron star as occuring through a two step process. Deconfinement
of nuclear matter to a two (up and down) flavour quark matter takes
place in the first step in strong interaction time scale. The second
step concerns with the generation of strange quarks from the excess
of down quarks via a weak process. We may add here that 
this is the first instance where a realistic nuclear matter EOS
is used to study the nuclear matter to SQM conversion as two step 
process. Drago {\it et al} \cite{key-20}, on the other hand, studied 
the burning of nuclear matter directly to SQM in detail by using the 
conservation conditions and the compact star models.

To study the conversion of nuclear matter to a two-flavour quark matter,
we here consider relativistic EOSs describing the forms of the matter
in respective phases. Along with such EOSs, we would also consider
hydrodynamical equations depicting various conservation conditions
to examine such conversion process in a compact neutron star. Development
of the conversion front, as it propagates radially through the model
star, would be examined. We would next study the conversion of two-flavour
quark matter to a three-flavour SQM through a non-leptonic weak interaction
process by assuming $\beta$ equilibrium for the SQM. The paper is
organised as follows. In section II, we discuss the EOSs used for
the present work. In section III, we discuss the conversion to two-flavour
quark matter. Conversion to three flavour SQM is discussed in section
IV. In section V, we summarise the results. Conclusions that may be
drawn from these results regarding the actual conversion process that
may take place in a neutron star are also presented in this final
section.

\section*{II. The equation of state}

The nuclear matter EOS has been evaluated using the nonlinear Walecka
model \cite{key-21}. The Lagrangian density in this model is given
by: 

\[
\mathcal{L}(x)=\sum_{i}\bar{\psi_{i}}(i\gamma^{\mu}\partial_{\mu}-m_{i}
+g_{\sigma i}\sigma+g_{\omega i}\omega_{\mu}\gamma^{\mu}
-g_{\rho i}\rho_{\mu}^{a}\gamma^{\mu}T_{a})\psi_{i}-
\frac{1}{4}\omega^{\mu\nu}\omega_{\mu\nu}\]

\[
+\frac{1}{2}m_{\omega}^{2}\omega_{\mu}\omega^{\mu}
+\frac{1}{2}(\partial_{\mu}\sigma\partial^{\mu}\sigma
-m_{\sigma}^{2}\sigma^{2})-\frac{1}{4}\rho_{\mu\nu}^{a}\rho_{a}^{\mu\nu}
+\frac{1}{2}m_{\rho}^{2}\rho_{\mu}^{a}\rho_{a}^{\mu}\]

\begin{equation}
-\frac{1}{3}bm_{n}(g_{\sigma N} {\sigma})^{3}-\frac{1}{4}C(g_{\sigma N} {\sigma})^{4}
+\bar{\psi_{e}}(i\gamma^{\mu}\partial_{\mu}-m_{e})\psi_{e}\qquad\qquad
\label{1}\end{equation}

The Lagrangian in equation (1) includes nucleons (neutrons and protons),
electrons, isoscalar scalar, isoscalar vector and isovector vector
mesons denoted by $\psi_{i}$, $\psi_{e}$, $\sigma$, $\omega^{\mu}$
and $\rho^{a,\mu}$, respectively. The Lagrangian also includes cubic
and quartic self interaction terms of the $\sigma$ field. The parameters
of the nonlinear Walecka model are meson-baryon coupling constants,
meson masses and the coefficient of the cubic and quartic self interaction
of the $\sigma$ mesons (b and c, respectively). The meson field interacts
with the baryons through linear coupling. The $\omega$ and $\rho$
meson masses have been chosen to be their physical masses. The rest
of the parameters, namely, nucleon-meson coupling constant 
($\frac{g_{\sigma}}{m_{\sigma}},\frac{g_{\omega}}{m_{\omega}}$
and $\frac{g_{\rho}}{m_{\rho}}$) and the coefficient
of cubic and quartic terms of the $\sigma$ meson self interaction
(b and c, respectively) are determined by fitting the nuclear matter
saturation properties, namely, the binding energy/nucleon (-16 MeV), baryon density
($\rho_{0}$=0.17 $fm^{-3}$), symmetry energy coefficient (32.5 MeV),
Landau mass (0.83 $m_{n}$) and nuclear matter incompressibility (300 MeV).

In the present paper, we first consider the conversion of nuclear
matter, consisting of only nucleons (\textit{i.e.} without hyperons)
to a two flavour quark matter. The final composition of the quark matter
is determined
from the nuclear matter EOS by enforcing the baryon number conservation
during the conversion process.
That is, for every neutron two down and one up quarks and for every
proton two up and one down quarks are produced, electron number being
same in the two phases. While describing the state of matter for the
quark phase we consider a range of values for the bag constant. Nuclear
matter EOS is calculated at zero temperature, whereas, the two flavour
quark matter EOS is obtained both at zero temperature as well as at
finite temperatures.

\section*{III. Conversion to two flavour matter}

In this section we discuss the conversion of neutron proton (n-p)
matter to two flavour quark matter, consisting of u and d quarks along
with electrons for the sake of ensuring charge neutrality. We heuristically
assume the existence of a combustion front. Using the macroscopic
conservation conditions, we examine the range of densities for which
such a combustion front exists. We next study the outward propagation
of this front through the model star by using the hydrodynamic (\textit{i.e.}
Euler) equation of motion and the equation of continuity for the energy
density flux \cite{key-23}. In this study, we consider a non-rotating,
spherically symmetric neutron star. The geometry of the problem effectively
reduces to a one dimensional geometry for which radial distance from the 
centre of the model star is the only independent spatial variable of interest.

Let us now consider the physical situation where a combustion front
has been generated in the core of the Neutron star. This front propagates
outwards through the neutron star with a certain hydrodynamic velocity, 
leaving behind a u-d-e matter. In the following, we denote all the physical
quantities in the hadronic sector by subscript 1 and those in the
quark sector by subscript 2.

Condition for the existence of a combustion front is given by \cite{key-23a}
\begin{equation}
\epsilon_{2}(p,X)<\epsilon_{1}(p,X),\label{1a}
\end{equation}
 where $p$ is the pressure and $X=(\epsilon+p)/{n_{B}}^{2}$, $n_{B}$
being the baryon density. Quantities on opposite sides of the front
are related through the energy density, the momentum density and the
baryon number density flux conservation. In the rest frame of the
combustion front, these conservation conditions can be written as
\cite{key-17,key-23,key-23a1}:

\begin{equation}
\omega_{1}v_{1}^{2}\gamma_{1}^{2}+p_{1}=\omega_{2}v_{2}^{2}\gamma_{2}^{2}+p_{2},
\label{2}\end{equation}

\begin{equation}
\omega_{1}v_{1}\gamma_{1}^{2}=\omega_{2}v_{2}\gamma_{2}^{2},
\label{3}\end{equation}
and
\begin{equation}
n_{1}v_{1}\gamma_{1}=n_{2}v_{2}\gamma_{2}.
\label{4}\end{equation}

In the above three conditions $v_{i}$ (i=1,2) is the velocity, $p_{i}$
is the pressure, $\gamma_{i}=\frac{1}{\sqrt{1-v_{i}^{2}}}$ is the
Lorentz factor, $\omega_{i}=\epsilon_{i}+p_{i}$ is the specific enthalpy
and $\epsilon_{i}$ is the energy density of the respective phases.

Besides the conservation conditions given in (\ref{1a}-\ref{4}), the
condition of entropy increase across the front puts an additional
constraint on the possibility of the existence of the combustion front.
This entropy condition is given by \cite{key-23b}, 
\begin{equation}
s_{1}v_{1}\gamma_{1}\leq s_{2}v_{2}\gamma_{2}\label{4a}
\end{equation}
 with $s_{i}$ being the entropy density.

The velocities of the matter in the two phases, given by equations
(\ref{2}-\ref{4}), are written as \cite{key-23}:

\begin{equation}
v_{1}^{2}=\frac{(p_{2}-p_{1})(\epsilon_{2}+p_{1})}{(\epsilon_{2}
-\epsilon_{1})(\epsilon_{1}+p_{2})},
\label{5}\end{equation}

and \begin{equation}
v_{2}^{2}=\frac{(p_{2}-p_{1})(\epsilon_{1}+p_{2})}{(\epsilon_{2}
-\epsilon_{1})(\epsilon_{2}+p_{1})}.
\label{6}\end{equation}

It is possible to classify the various conversion mechanisms by comparing
the velocities of the respective phases with the corresponding velocities
of sound, denoted by $c_{si}$, in these phases. Thes conditions are
\cite{key-23c},

\begin{center}strong detonation : $v_{1}>c_{s1},\qquad v_{2}<c_{s2}$, \end{center}

\begin{center}Jouget detonation : $v_{1}>c_{s1},\qquad v_{2}=c_{s2}$, \end{center}

\begin{center}supersonic or weak detonation: $v_{1}>c_{s1},\qquad v_{2}>c_{s2}$, \end{center}

\begin{center}strong deflagration : $v_{1}<c_{s1},\qquad v_{2}>c_{s2}$, \end{center}

\begin{center}Jouget deflagration : $v_{1}<c_{s1},\qquad v_{2}=c_{s2}$, \end{center}

\begin{center}weak deflagration : $v_{1}<c_{s1},\qquad v_{2}<c_{s2}$. \end{center}

For the conversion to be physically possible, velocities should satisfy
an additional condition, namely, $0\leq v_{i}^{2}\leq 1$. We here
find that the velocity condition, along with the eq.(\ref{1a}),
puts severe constraint on the allowed equations of state.

To examine the nature of the hydrodynamical front, arising from the
neutron to two flavour quark matter conversion, we plot, in fig.1,
the quantities $v_{1},v_{2},c_{s1}$ and $c_{s2}$ as functions of
the baryon number density ($n_{B}$). As mentioned earlier, the u
and d quark content in the quark phase is kept same as the one corresponding
to the quark content of the nucleons in the hadronic phase. With these
fixed densities of u and d quarks and electrons, the EOS of the two
flavour matter has been evaluated using the bag model prescription.
We find that the energy condition (eqn.(\ref{1a})) and velocity
condition ( ${v_{i}}^{2}>0$ ) both are satisfied only for a small
window of $\approx\pm5.0MeV$ around the bag pressure $B^{1/4}=160MeV$.
The constraint imposed by the above conditions results in the possibility
of deflagration, detonation or supersonic front as shown in the 
figs.(1-3).

In fig.1, we considered both the phases to be at zero temperature.
A possibility, however, exist that a part of the internal energy is
converted to heat energy, thereby increasing the temperature of the
two flavour quark matter during the exothermic combustive conversion
process. Instead of following the prescription for the estimation
of temperature as given in references \cite{key-15,key-20}, we study
the changes in the properties of combustion with the temperature of
the newly formed two flavour quark phase in the present paper. In
figures 2 and 3, we plot the variation of velocities with density
at two different temperatures, namely, $T=50MeV$ and $T=100MeV$,
respectively. These figures show that the range of values of baryon
density, for which the flow velocities are physical, increases with
temperature. Figure 4 shows the variation of velocities with temperature
for values of baryon number densities given by $n_{B} \approx 3\rho_{0}$
and $7\rho_{0}$, respectively. In this figure, the difference between
velocities $v_{1}$ and $v_{2}$ increases with temperature of the
two flavour quark matter. In the present paper we have considered only
the zero temperature nuclear matter EOS. 
On the other hand, equation of state of quark matter
has a finite temperature dependence and hence the difference between
${v_{1}}$ and ${v_{2}}$, varies with temperature.

The preceding discussion is mainly a feasibility study for the possible
generation of the combustive phase transition front and its mode of
propagation. Having explored such possibilities, we now study the
evolution of the hydrodynamical combustion front with position as
well as time. This might give us some insight regarding the actual
conversion of a hadronic star to a quark star and the time scale involved
in such a process. To examine such an evolution, we move to a reference
frame in which the nuclear matter is at rest. The speed of the combustion
front in such a frame is given by ${v}_{f}={-v}_{1}$ with $v_{1}$
being the velocity of the nuclear matter in the rest frame of the
front.

In the present work, we use the special relativistic formalism to
study the evolution of combustion front as it moves outward in the
radial direction inside the model neutron star. The relevant equations
are the equation of continuity and the Euler's equation, that are
given by \cite{key-17}:

\begin{equation}
\frac{1}{\omega}(\frac{\partial\epsilon}{\partial\tau}
+v\frac{\partial\epsilon}{\partial r})+
\frac{1}{W^{2}}(\frac{\partial v}{\partial r}
+v\frac{\partial v}{\partial\tau})+2\frac{v}{r}=0
\label{7}\end{equation}

and 
\begin{equation}
\frac{1}{\omega}(\frac{\partial p}{\partial r}+
v\frac{\partial p}{\partial\tau})+
\frac{1}{W^{2}}(\frac{\partial v}{\partial\tau}+
v\frac{\partial v}{\partial r})=0,
\label{8}\end{equation}
 where, $v=\frac{\partial r}{\partial\tau}$ is the front velocity
in the nuclear matter rest frame and $k=\frac{\partial p}{\partial\epsilon}$
is taken as the square of the effective sound speed in the medium.

Substituting these expressions for $v$ and $k$ in equations (\ref{7})
and (\ref{8}) we get 
\begin{equation}
\frac{2v}{\omega}\frac{\partial\epsilon}{\partial r}
+\frac{1}{W^{2}}\frac{\partial v}{\partial r}(1+v^{2})+
\frac{2v}{r}=0
\label{9}\end{equation}

and 
\begin{equation}
\frac{n}{\omega}\frac{\partial\epsilon}{\partial r}(1+v^{2})
+\frac{2v}{W^{2}}\frac{\partial v}{\partial r}=0
\label{10}\end{equation}

Equations (\ref{9}) and (\ref{10}) ultimately yield single differential
equation that is written as:

\begin{equation}
\frac{dv}{dr}=\frac{2vkW^{2}(1+v^{2})}{r[4v^{2}-k(1+v^{2})^{2}]}.
\label{11}\end{equation}

The equation (\ref{11}) is integrated, with respect to r(t), starting
from the centre towards the surface of the star. The nuclear and quark
matter EOS have been used to construct the static configuration of
compact star, for different central densities, by using the standard
Tolman-Oppenheimer-Volkoff (TOV) equations \cite{key-17a}. The velocity
at the centre of the star should be zero from symmetry considerations.
On the other hand, the 1/r dependence of the dv/dr, in eq.(\ref{11})
suggests a steep rise in velocity near the centre of the star. 

Our calculation proceeds as follows. We first construct the density
profile of the star for a fixed central density. Equations (\ref{5})
and (\ref{6}) then specify the respective flow velocities $v_{1}$
and $v_{2}$ of the nuclear and quark matter in the rest frame of
the front, at a radius infinitesimally close to the centre of the
star. This would give us the initial velocity of the front ($-v_{1}$),
at that radius, in the nuclear matter rest frame. We next start with
equation (\ref{11}) from a point infinitesimally close to the centre
of the star and integrate it outwards along the radius of the star.
The solution gives us the variation of the velocity with the position
as a function of time of arrival of the front, along the radius of
the star. Using this velocity profile, we can calculate the time required
to convert the whole star using the relation $v=dr/d\tau$.

In figure 5, we show the variation of the velocity for values of the
central baryon densities 3$\rho_{0}$, 4.5$\rho_{0}$ and 7$\rho_{0}$,
respectively. The respective initial velocities corresponding to such
central densities are taken to be 0.43, 0.67 and 0.68. The figure
shows that the velocity of the front, for all the central densities,
shoots up near the centre and then saturates at a certain velocity
for higher radius. Such a behaviour of velocity near the central point
is apparent from the equation (\ref{11}) above. The numerically obtained
saturation velocity varies from 0.92 for central baryon density 3$\rho_{0}$
to 0.98 for 7$\rho_{0}$. The existence of a saturation velocity,
at large r, is apparent from the asymptotic behaviour of equation
(\ref{11}). A comparison with fig.1 shows that for the densities
3$\rho_{0}$ and 4.5$\rho_{0}$, the conversion starts as weak detonation
and stays in the same mode throughout the star. On the other hand,
for 7$\rho_{0}$, initial detonation front changes over to weak detonation
and the velocity of front becomes almost 1 as it reaches the outer
crust. The corresponding time taken by the combustion front to propagate
inside the star is plotted against the radius in figure 6. The time
taken by the front to travel the full length of the star is of the
order of few milliseconds. According to the present model, the initial
neutron star thus becomes a two flavour quark star in about $10^{-3}$
sec. The results discussed above correspond to the case in which both
nuclear as well as quark matter are at zero temperature. For finite
temperature quark matter results vary only by a few percent of the
front velocities for the quark matter at zero temperature.

We would like to mention here that in the above discussions, the equations
governing the conversion of nuclear to quark matter are purely hydrodynamic.
There is no dissipative process, so that the combustion front continues
to move with a finite velocity depending on the density profile. Furthermore, 
there is no reaction rate involved here as the deconfinement process occurs 
in the strong interaction time scale and hence can be taken to be instantaneous.
This is certainly very different from the second step process, to
be discussed in the next section, where the two flavour matter converts
to a three flavour matter. Here, the governing rate equations are
weak interaction rates which play a decisive role in the conversion.
Comparing the total time ( $\equiv10^{-3}$ sec) taken by the combustion
front to travel through the star with the weak interaction time scale
($10^{-7}-10^{-8}$ sec), it is evident that the second step may start
before the end of the first step process. In that case, perhaps, one
should ideally considers two fronts, separated by a finite distance,
moving inside the star \cite{key-23}. In the present paper, we have
taken a much simplified picture and considered the conversion of a
chemically equilibrated two flavour to three flavour quark matter
as the second step process. Our results may provide us with more information
regarding the necessity of considering two fronts.

\section*{IV. The conversion to three flavour SQM}

In this section we discuss the conversion of two flavour quark matter
to three flavour SQM in a compact star. Similar to the discussion
above, we assume the existence of a conversion front at the core of
the star that propagates radially outward leaving behind the SQM as
the combustion product. This conversion is governed by weak interactions
that take place inside the star.

For a three flavour quark matter, the charge neutrality and the baryon
number conservation conditions yield

\begin{equation}
3n_{B}=n_{u}+n_{d}+n_{s}\label{12}
\end{equation}

\begin{equation}
2n_{u}=n_{d}+n_{s}+3n_{e}\label{13}
\end{equation}

where $n_{i}$ is the number density of particle i (i= u, d, s and
e).

The weak reactions which govern the conversion of excess down quark
to strange quark can be written as,

\begin{equation}
d\rightarrow u+e^{-}+\overline{\nu}_{e-};\;~~~s\rightarrow u+e^{-}+
\overline{\nu}_{e-};\;~~~d+u\rightarrow s+u
\end{equation}

We assume that the neutrinos escape freely from the site of reaction
and the temperature of the star remains constant. The non-leptonic
weak interaction in such a case becomes the governing rate equation. The semi-leptonic
weak decays, then, are solely responsible for the chemical equilibration
which can be incorporated through the relations given below.

\begin{equation}
\mu_{e^{-}}=\mu_{d}-\mu_{u};\;~~~\mu_{d}=\mu_{s},\label{14}
\end{equation}

where $\mu_{i}$ is the chemical potential of the i'th particle. The
number densities ($n_{i}$) of the quarks and electrons are related
to their respective chemical potentials by,

\begin{equation}
n_{i}=g_{i}\int_{0}^{\infty}d^{3}p/(2\pi)^{3}[f^{+}-f^{-}],\label{15}
\end{equation}

where $f^{+}$and $f^{-}$ are given by

\begin{equation}
f^{+}=\frac{1}{exp[(E_{p}-\mu)/T]+1}\qquad \qquad f^{-}=\frac{1}{exp[(E_{p}+\mu)/T]+1}.
\label{16}
\end{equation}
In equations (\ref{15}) and (\ref{16}), $g_{i}$ is the degeneracy
factor and $T$ the temperature. Eqns.(\ref{12}-\ref{16}) can be
solved self consistently to calculate the number densities of quarks
and electrons.

The conversion to SQM starts at the centre ($r=0$) of the two flavour
star. Assuming that the reaction region is much smaller than the size
of the star, we have considered the front to be one dimensional. Moreover,
as we are considering spherical static stars only, there is no angular
dependence. The combustion front, therefore, moves radially towards
the surface of the star. As the front moves outwards, excess d quarks
get converted to s quark through the non leptonic weak process. The
procedure employed in the present work is somehow similar to that
of ref.\cite{key-9}, although the physical boundary conditions are
different.

We now define a quantity,

\begin{equation}
a(r)=[n_{d}(r)-n_{s}(r)]/2n_{B}\label{17}
\end{equation}
such that, $a(r=0)=a_{0}$ at the core of the star. The quantity
$a_{0}$ is the number density of the strange quarks, at the
centre, for which the SQM is stable and its value lies between 0 and
1. For equal numbers of d and s quarks, $a(r)=0$. Ideally at the
centre of the star $a_{0}$ should be zero for strange quark mass
$m_{s}=0$. Since s quark has a mass $m_{s}\sim150$MeV, at the centre
of the star $a_{0}$ would be a small finite number, depending on
the EOS. The s quark density fraction, however, decreases along with
the decrease of the baryon density towards the surface of the star,
so that, $a(r\rightarrow R)\rightarrow1$ with R being the radius
of the star. At any point along the radius, say $r=r_{1}$, initial
$a(r_{1})$, before the arrival of the front, is decided by the initial
two flavour quark matter EOS. The final $a(r_{1})$, after the conversion
is obtained from the equilibrium SQM EOS at the density corresponding
to $r_{1}$. 

The conversion to SQM occurs via decay of down quark to strange quark
($u+d\rightarrow s+u$) and the diffusion of the strange quark
from across the front \cite{key-9}. The corresponding rate of change
of $a(r)$ with time is governed by following two equations: \begin{equation}
\frac{da}{dt}=-R(a),\label{18}\end{equation}
 and \begin{equation}
\frac{da}{dt}=D\frac{d^{2}a}{dr^{2}},\label{19}\end{equation}
 In the equation (\ref{18}) $R(a)$ is the rate of conversion of
d to s quarks. Equation (\ref{19}) yields the rate of change of $a(r)$
due to diffusion of s quarks, with $D$ being the diffusion constant.
Following Olinto \cite{key-9}, assuming the one dimensional steady
state solution and using equations (\ref{18}) and \ref{19} we get:
\begin{equation}
Da^{''}-va^{'}-R(a)=0,\label{20}
\end{equation}
where $v$ is the velocity of the fluid. In equation (\ref{20})
$a^{'}=\frac{da}{dr}$.

Conservation of baryon number flux at any position yields $n_{q}v_{q}=n_{s}v_{s}$.
The subscripts q and s denote the two flavour quark matter phase and
SQM phase, respectively. The baryon flux conservation condition yields
the initial boundary condition at any point $r$ along the radius
of the star: 
\begin{equation}
a'(r)=-\frac{v}{D}({a_{i}}(r)-{a_{f}}(r)),\label{20a}
\end{equation}
where ${a_{i}}(r)$ and ${a_{f}}(r)$ are the values for the $a(r)$
before and after the combustion, respectively.

The reaction rate for the non-leptonic weak interaction $u+d\rightarrow u+s$
is in general a five dimensional integral for non zero temperature
and $m_{s}$ \cite{key-25,madsen}. Here, instead, we have taken the
zero temperature, small $a$ limit \cite{key-9}. 
\begin{equation}
R(a)\approx\frac{16}{15\pi}{G_{F}}^{2}{cos^{2}}\theta_{c}{sin^{2}}\theta_{c}{\mu_{u}}^{5}
{\frac{a}{3}}^{3},\label{20b}
\end{equation}
where, $G_{F}$ is the weak coupling constant and $\theta_{c}$ is
the Cabibbo angle. The above equation can be written in the following
form: 
\begin{equation}
R(a)\approx\frac{a^{3}}{\tau},\label{20c}
\end{equation}
where $\tau=\frac{16}{(3^{3})15\pi}{G_{F}}^{2}{cos^{2}}\theta_{c}{sin^{2}}\theta_{c}{\mu_{u}}^{5}$,
here, depends on the position of the front.

Following the line of arguments given in ref.\cite{key-9}, we write
down the analytic expressions for $D$ and $v$ as:

\begin{equation}
D=\frac{\lambda\overline{v}}{3}\simeq10^{-3}(\frac{\mu}{T})^{2}cm^{2}/s,\label{22}
\end{equation}
and 
\begin{equation}
v=\sqrt{{\frac{D}{\tau}\frac{{{a_{f}}(r)}^{4}}{2({a_{i}}(r)-{a_{f}}(r))}}}.\label{23}
\end{equation}

Our calculation proceeds as follows. First we get the star characteristics
for a fixed central baryon density $\rho_{c}$. For a given $\rho_{c}$,
number densities of u, d and s quarks, in both the two and three flavour
sectors, are known at any point inside the star. That means ${a_{i}}(r)$
and ${a_{f}}(r)$ is fixed. Eqns.(\ref{20b}-\ref{23}) are then used
to get the diffusion constant and hence the radial velocity of the
front.

The central baryon densities considered here are same as those of
section III. Assuming that the neutrinos leave the star, the temperature
is kept constant at some small temperature so that we can use the
equation (\ref{20b}), evaluated in the zero temperature limit. The
variation of $a(r)$ with the radius of the star is given in figure
7. The plot shows that $a(r)$ increases radially outward, which corresponds
to the fact that as density decreases radially, the number of excess
down quark which is being converted to strange quark by weak interaction
also decreases. Hence, it takes less time to reach a stable configuration
and hence the front moves faster, as shown in the figs. 8 and 9.

In Fig.8, we have plotted the variation of velocity along the radius
of the star. The velocity shows an increase as it reaches sufficiently
low density and then drops to zero near the surface as $d\rightarrow s$
conversion rate becomes zero. Fig.9 shows the variation of time taken
to reach a stable configuration at different radial position of the
star. The total time needed for the conversion of the star, for different
central densities, is of the order of 100 seconds, as can be seen
from the figure.

\section*{V. Summary and discussion}

We have studied the conversion of a neutron star to strange star.
This conversion takes place in two stages. In the first stage a detonation
wave is developed in the hadronic matter (containing neutrons, protons
and electrons). We have described this hadronic matter with a relativistic
model. For such an equation of state the density profile of the star
is obtained by solving Tolman-Openheimer-Volkoff equations. The corresponding
quark matter equation of state is obtained by using the bag model.
However, this quark matter equation of state is not equilibrated and
contains two flavours. Matter velocities in the two media, as measured
in the rest frame of the front, have been obtained using conservation
conditions. These velocities have been compared with the sound velocity
in both phases.

For a particular density inside the star, flow velocities of the matter
on the two sides of the front is now fixed. Starting from a point,
infinitesimally close to the centre, hydrodynamic equations are solved
radially outward. The solution of the hydrodynamic equations gives
the velocity profiles for different central densities. The velocity
of the front shoots up very near to the core and then saturates at
a value close to 1. The mode of combustion is found to be weak detonation for 
lower central densities. For higher central densities, the initial 
detonation becomes weak detonation as the the front moves radially 
outward inside the star. This result is different from that of 
ref.\cite{key-20},
where the conversion process always correspond to a strong deflagration.
The time required for the conversion of the neutron
star to a two flavour quark star is found to be of the order of few
milliseconds. After this front passes through, leaving behind a two
flavour matter a second front is generated. This second front converts
the two flavour matter via weak interaction processes. The velocity
of the front varies along the radius of the star. As the front moves
out from the core to the crust, its velocity increases, implying faster
conversion. The time for the second conversion to take place comes
out to be $\sim100$ seconds. This is comparable to the time scale
obtained in ref.\cite{key-9}.

The comparison of the time of conversion from neutron star to two
flavour quark star and the weak interaction time scale suggests that
at some time during the passage of the first combustion front, the
burning of two flavour matter to strange matter should start. This
means that at some point of time, there should be two fronts moving
inside the star. On the other hand, our results show that, inside
the model star, the burning of the nuclear matter to two flavour quark
matter takes much smaller time compared to the conversion from two
flavour quark matter to SQM. However, the consideration of two fronts
might provide us with some more information regarding the conversion
of neutron star to a final stable strange star. In the present case
we have considered a two step process, there being only one type of
front, inside the star, at any instant of time.

Here we would also like to mention that ideally the second step should
start with a non-equilibrated two flavour matter \cite{key-24,key-25}.
Since this is a numerically involved calculation, in the present case
we have taken the simplified picture of equilibrated quark matter.

Finally, the burning of the nuclear matter to two flavour quark matter
is studies using special relativistic hydrodynamic equations. The
actual calculation should involve general relativity, taking into
account the curvature of the front for the spherical star. We propose
to explore all these detailed features in our subsequent papers.

\section*{Acknowledgements}

R.M. would like to thank CSIR for financial support. S.K.G., S.R. and P.S.J.
,in particular, thanks DST for financial support under IRHPA scheme.

\newpage
\begin{figure}
\includegraphics[%
  width=4in]{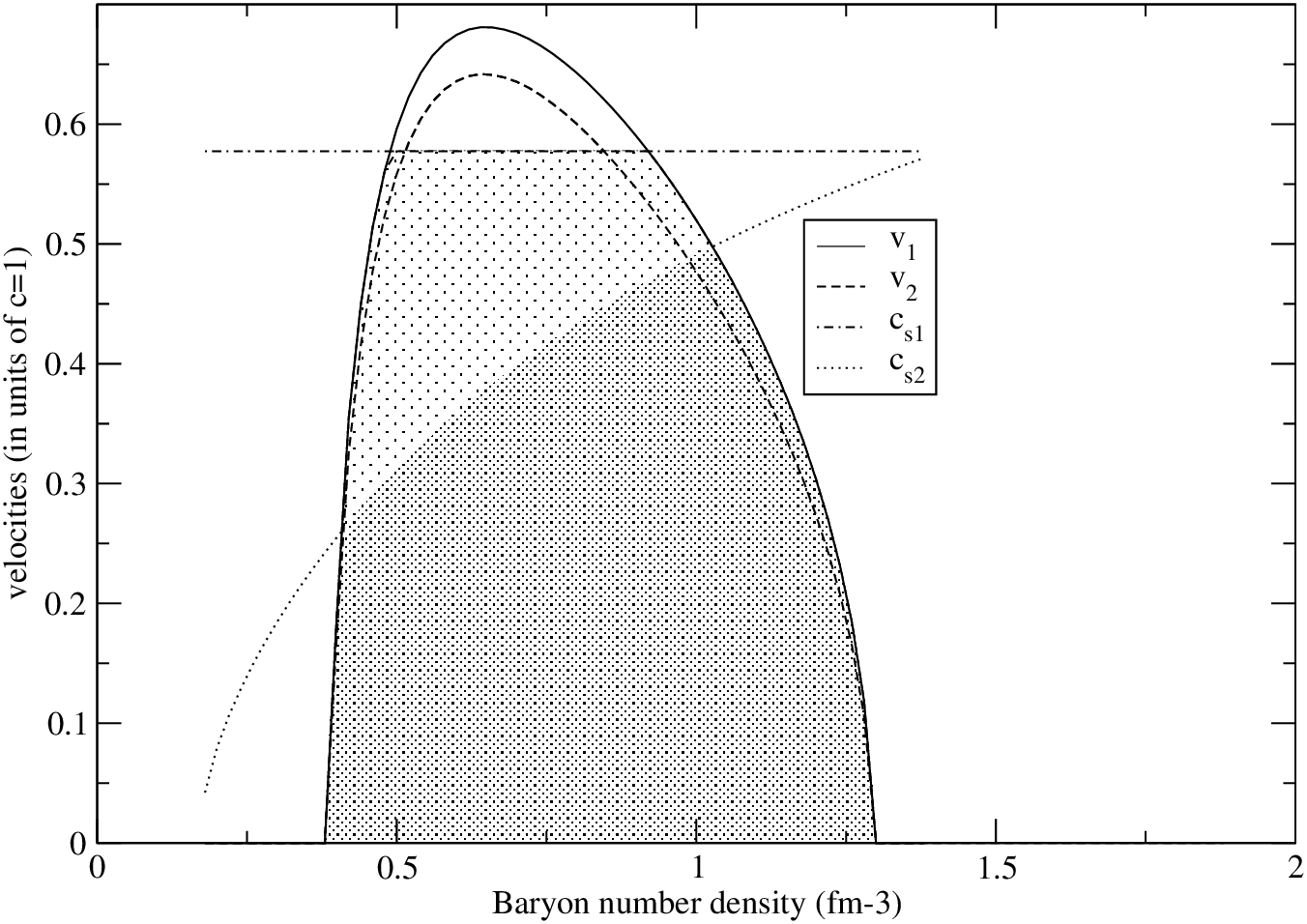}

\caption{Variation of different velocities with baryon number density for
T = 0 MeV. The dark-shaded region correspond to deflagration, light-shaded
region correspond to detonation and the unshaded region
correspond to supersonic conversion processes.}
\end{figure}

\begin{figure}
\includegraphics[%
  width=4in]{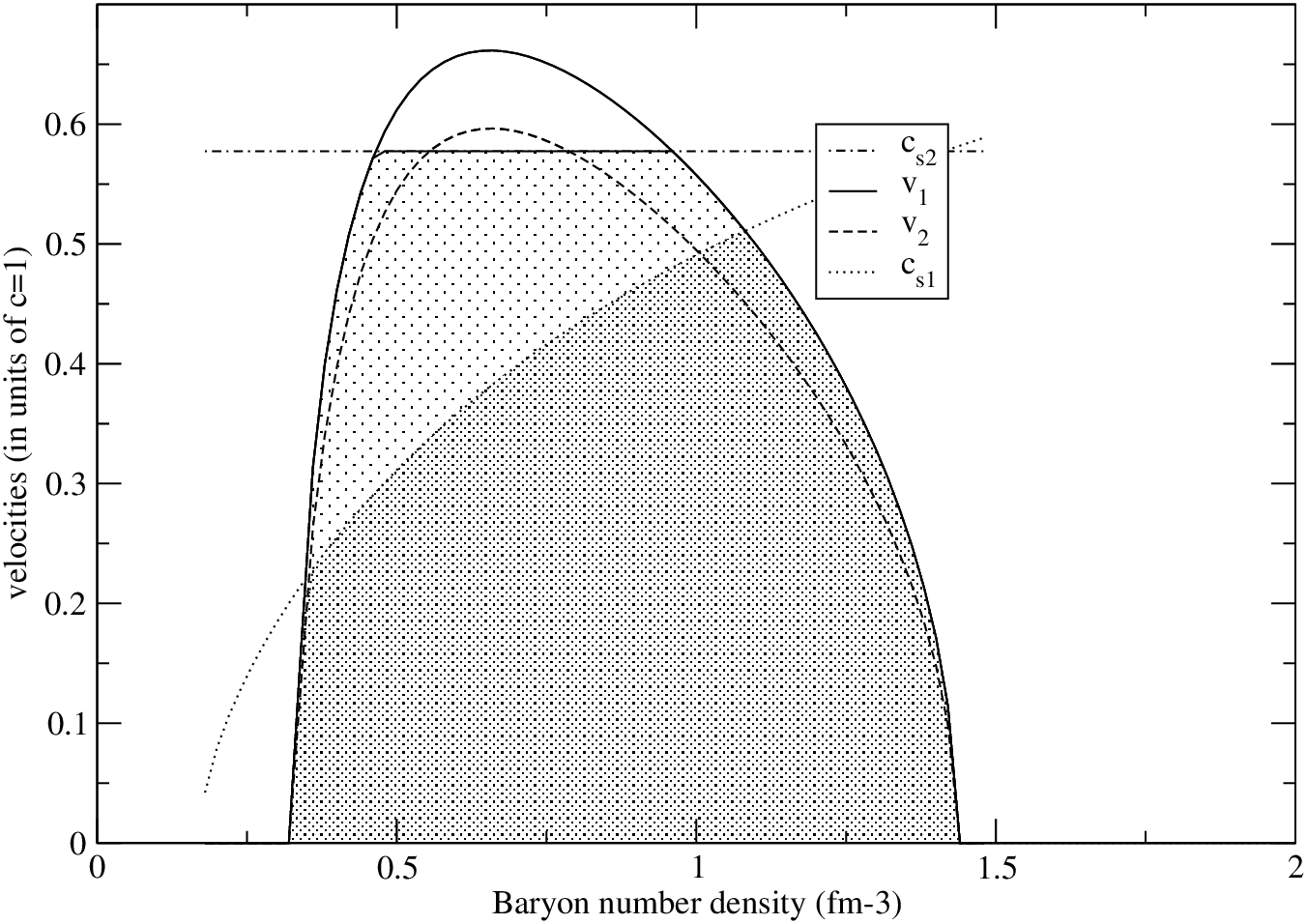}

\caption{Variation of velocities with baryon number density for T = 50 MeV.
Different regions correspond to different modes of conversion, 
where the notations are the same as in fig 1.}
\end{figure}

\begin{figure}
\includegraphics[%
  width=4in]{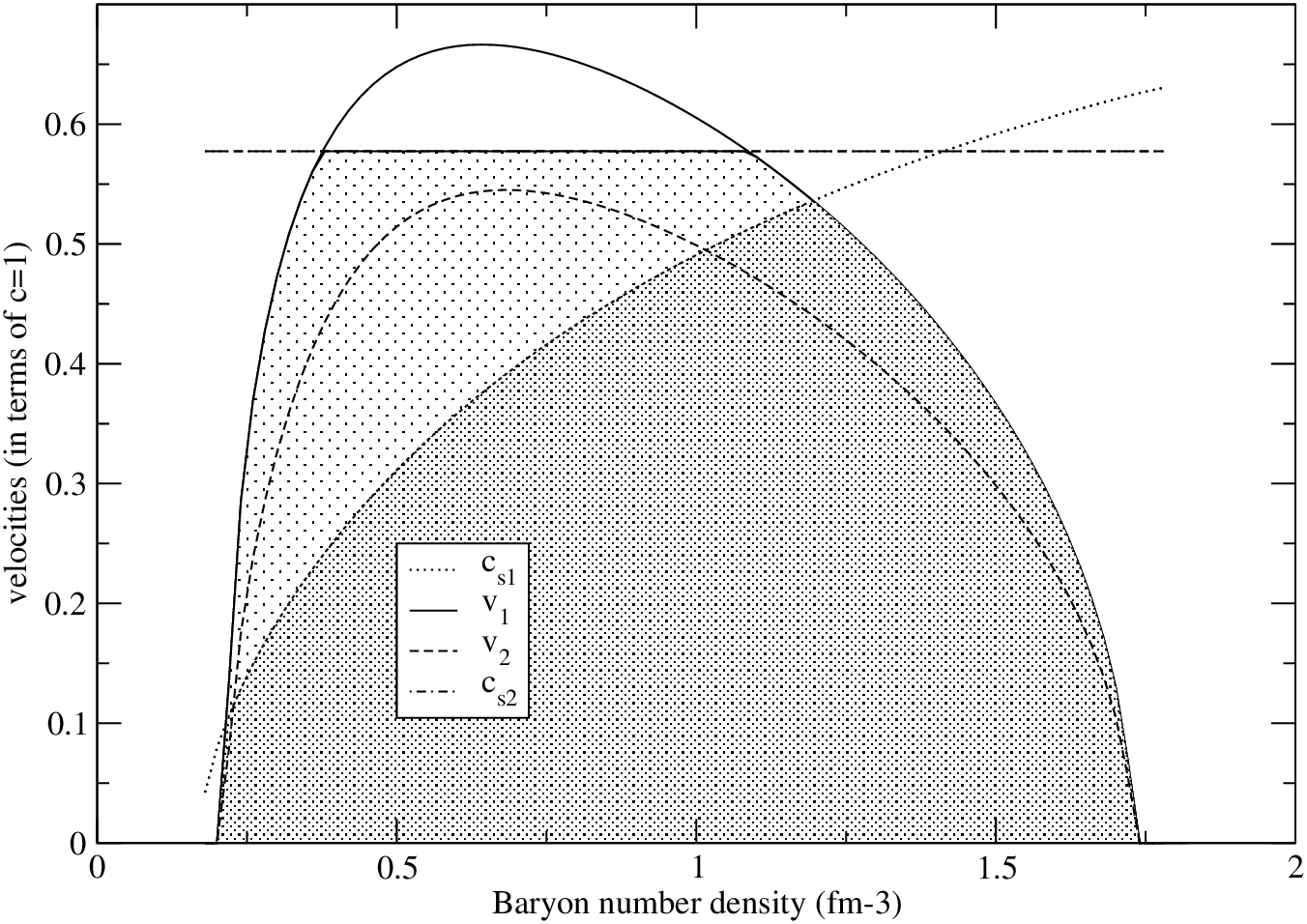}

\caption{Variation of velocities with baryon number density for T = 100 MeV.
Different regions correspond to different modes of conversion, 
where the notations are the same as in fig 1.}
\end{figure}

\begin{figure}
\includegraphics[%
  width=4in]{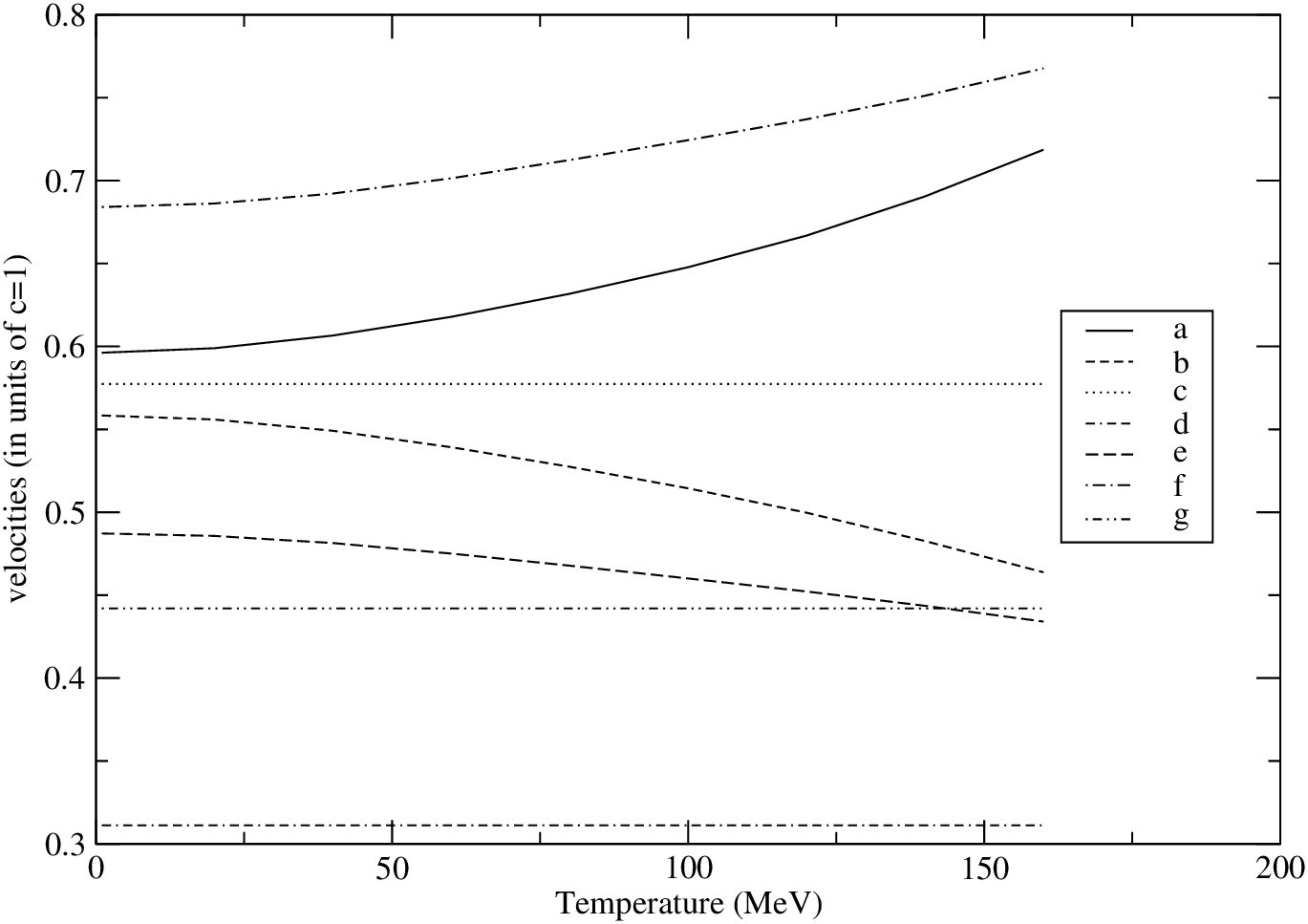}

\caption{Variation of velocities with temperature: (a)\( v_1 \) for density
3\(\rho_0 \), (b) \( v_2 \) for density 3\( \rho_0 \), (c) \( c_{s1} \),
(d) \( c_{s2} \) for 3\( \rho_0 \), (e) \( v_2 \) for 7\( \rho_0 \),
(f) \( v_1 \) for 7\( \rho_0 \) and (g) \(c_{s2} \) for 7\( \rho_0 \).}
\end{figure}

\begin{figure}
\includegraphics[%
  width=4in]{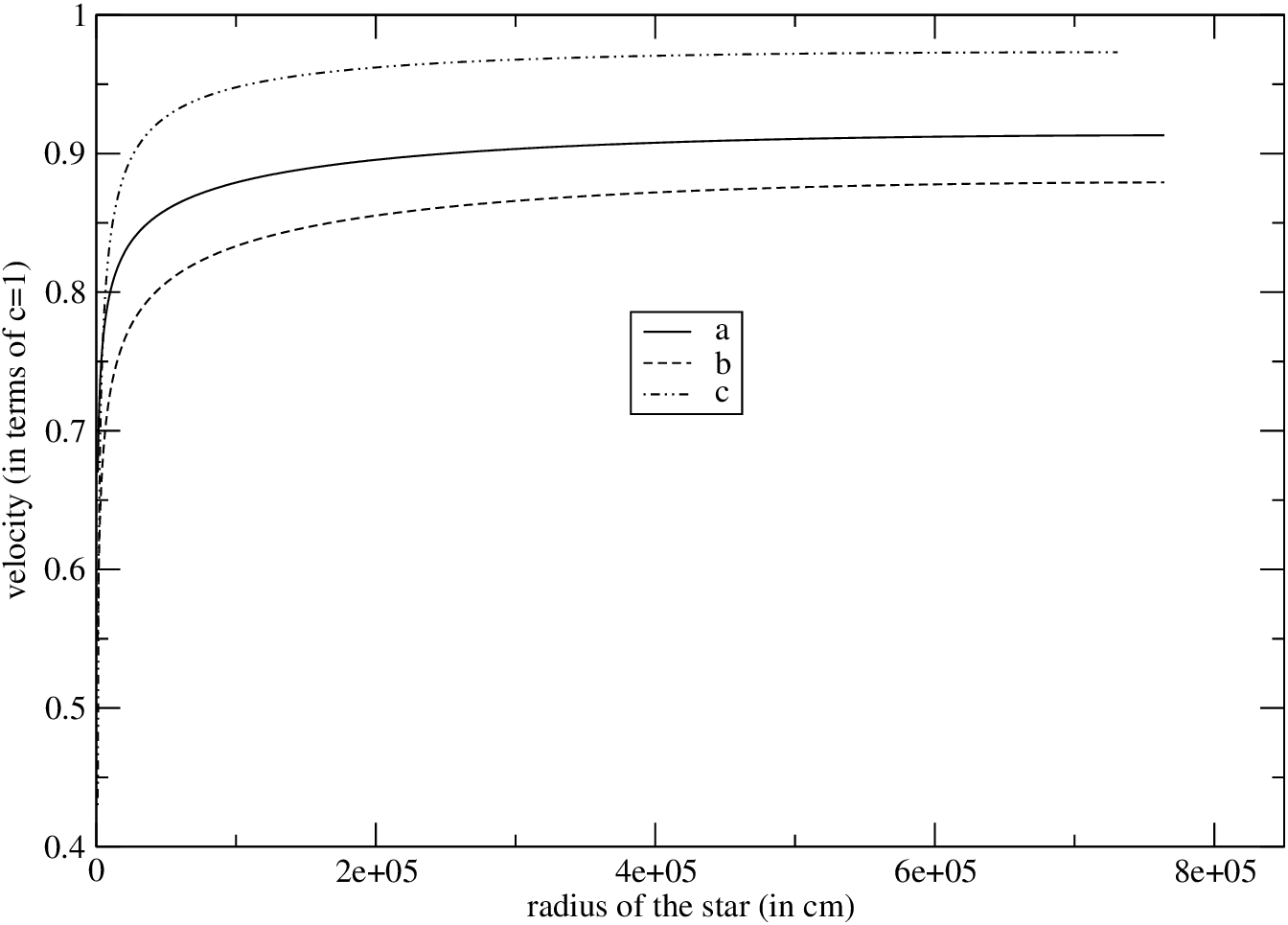}

\caption{Variation of velocity of the conversion front with radius of the
star, for three different values of the central densities, namely,
(a) $3\rho_{0}$, (b) $4.5\rho_{0}$ and (c) $7\rho_{0}$, respectively.
Here $\rho_{0}$ is the nuclear density. The initial velocity for
the three cases are 0.66, 0.65 and 0.47, respectively.}
\end{figure}

\begin{figure}
\includegraphics[%
  width=4in]{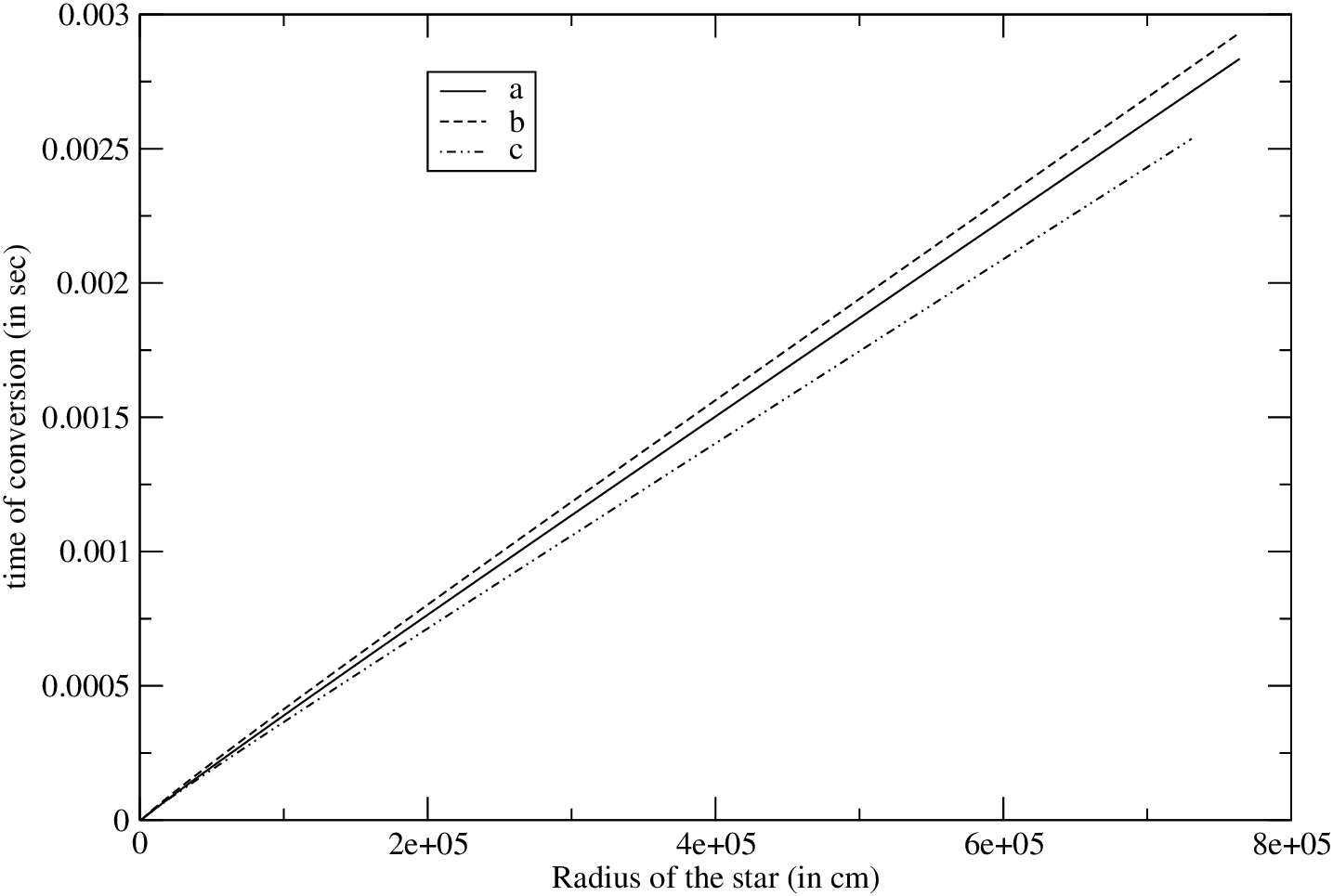}

\caption{Variation of the time of arrival of the conversion front at a certain
radial distance inside the star as a function of that radial distance from the centre
of the star for three different central densities. Values of the central
density corresponding to the curves (a), (b) and (c) are the same
as shown in fig 5.}
\end{figure}

\begin{figure}
\includegraphics[%
  width=4in]{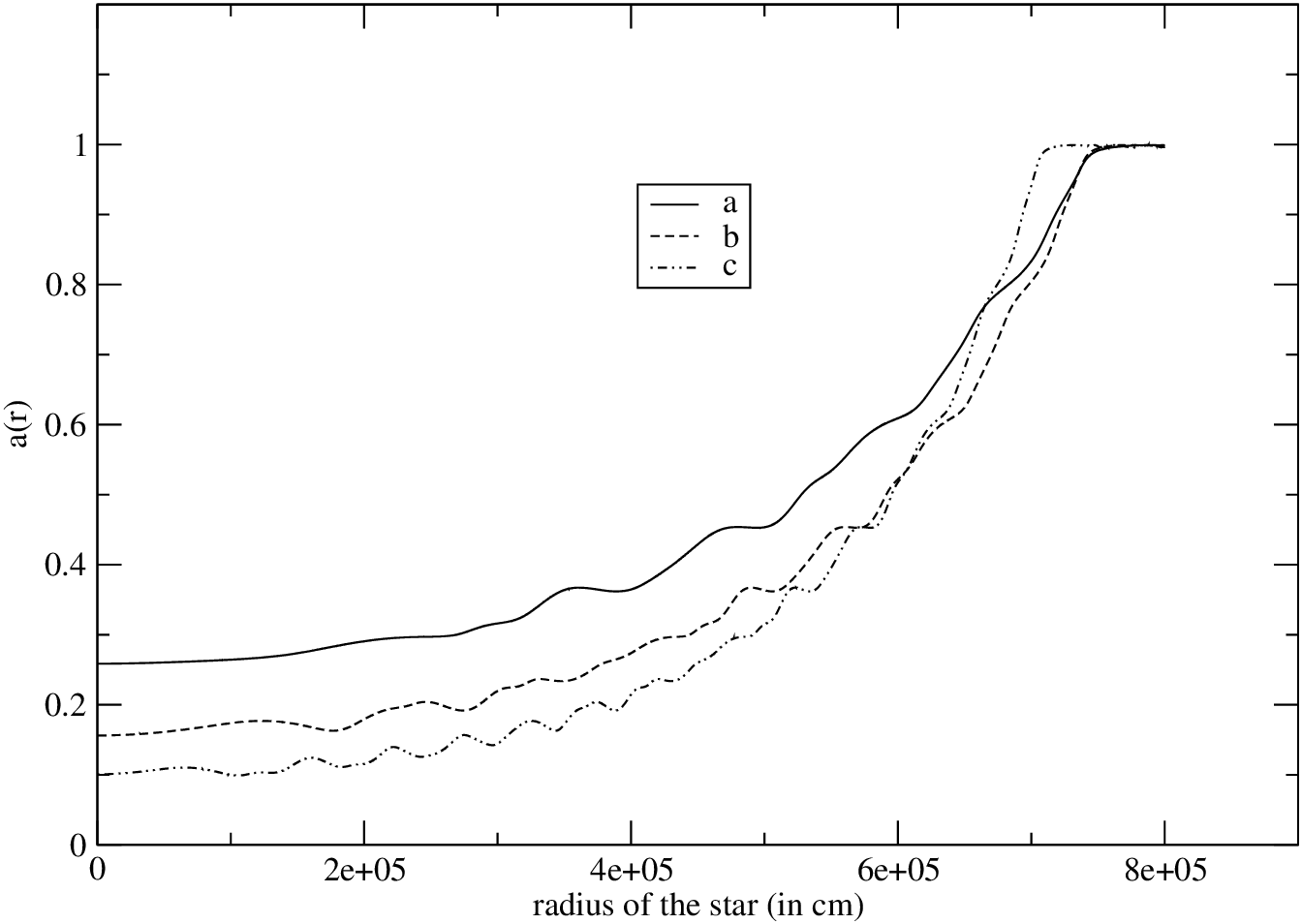}

\caption{Variation of a(r) (as in the text) with radius of star,
for different central densities, where, (a) corresponds to case for
which the central density is $3\rho_{0}$, (b) for $4.5\rho_{0}$
and (c) for $7\rho_{0}$.}
\end{figure}

\begin{figure}
\includegraphics[%
  width=4in]{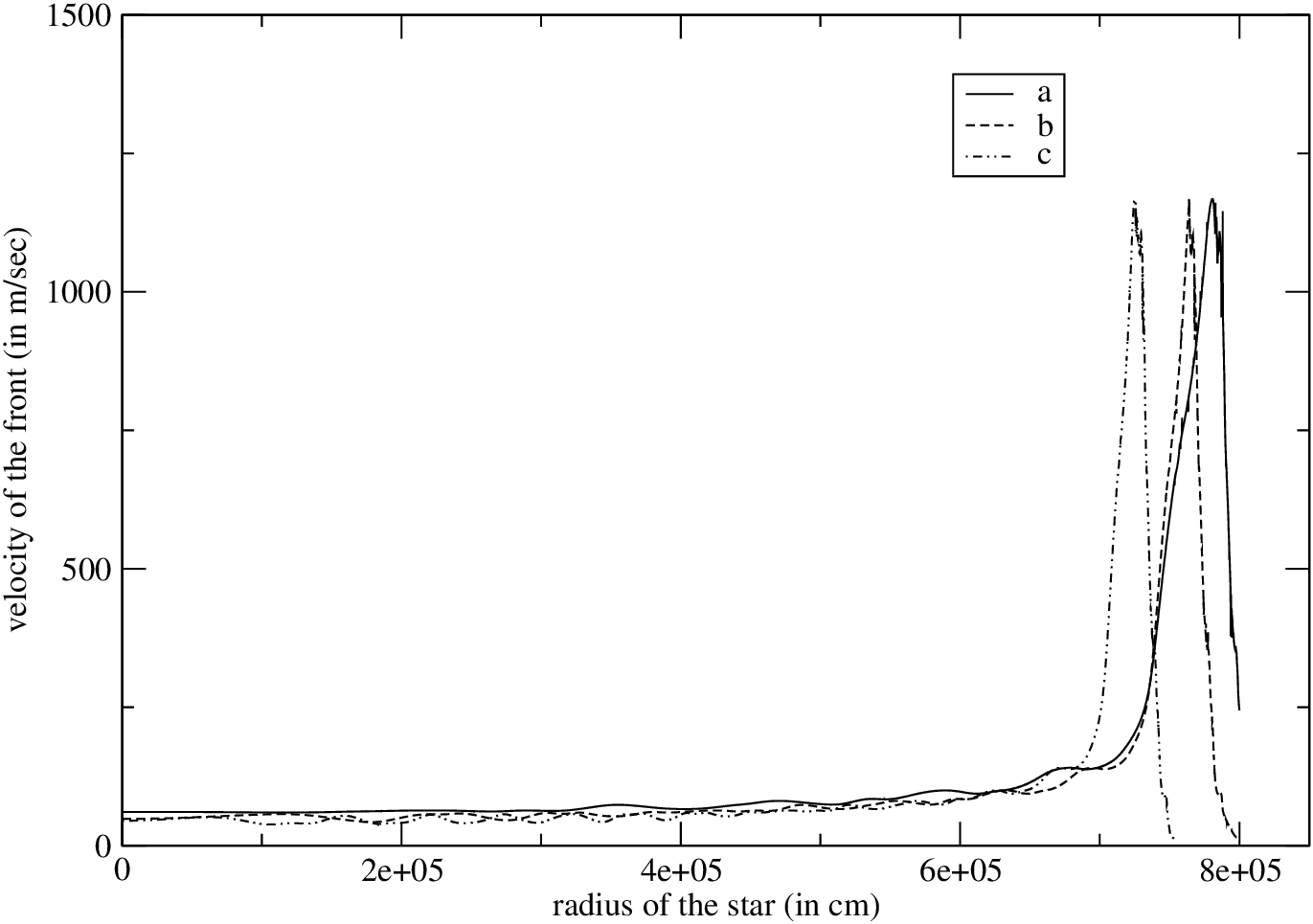}

\caption{Variation of velocity of the two to three flavour quark conversion
front with radius of the star for different central densities, where,
(a) corresponds to case for which the central density is $3\rho_{0}$,
(b) for $4.5\rho_{0}$ and (c) for $7\rho_{0}$. }
\end{figure}

\begin{figure}
\includegraphics[%
  width=4in]{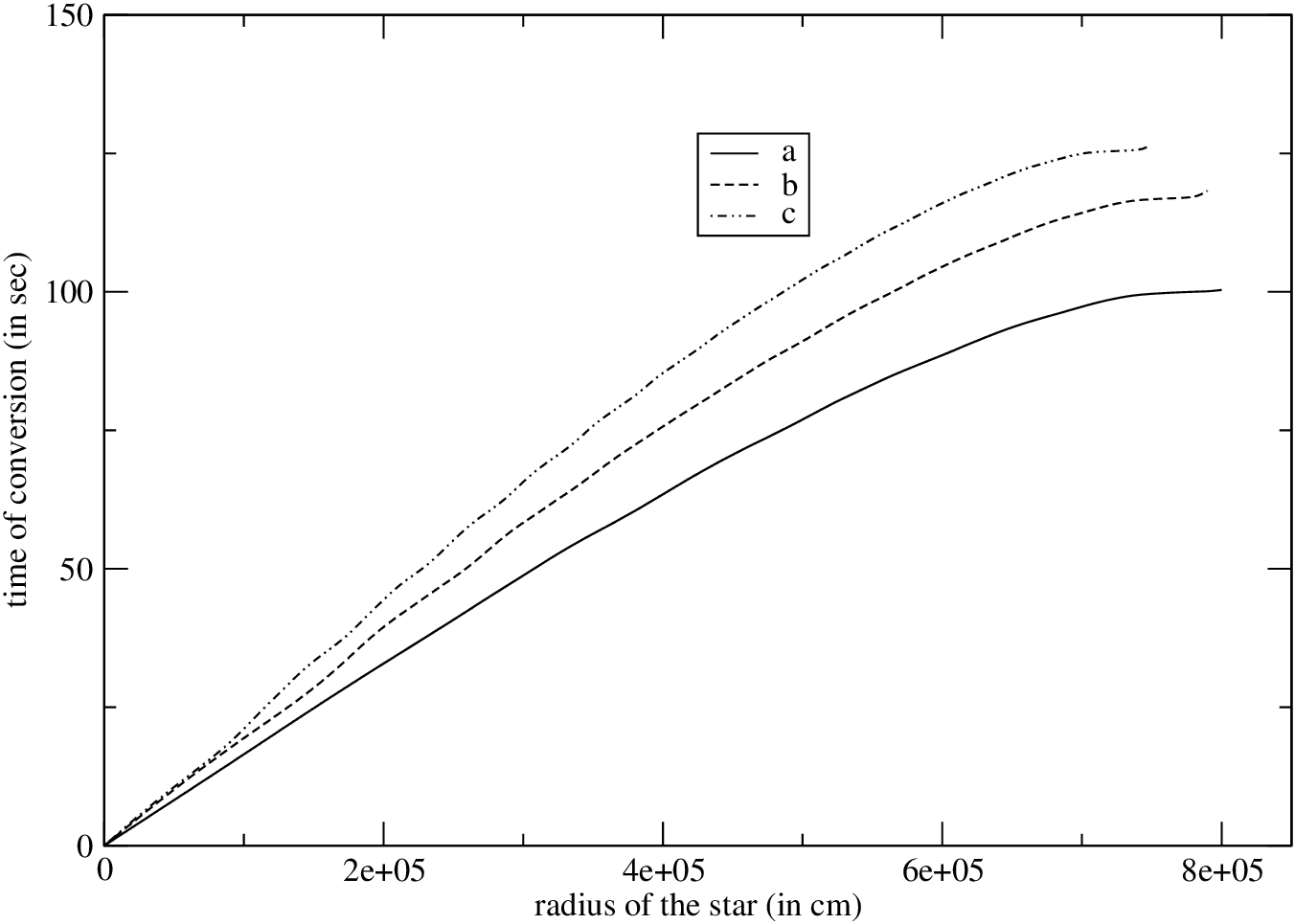}

\caption{Variation of time taken for the two to three flavour quark conversion
front with radius of the star, for different central densities, where,
(a) corresponds to case for which the central density is $3\rho_{0}$,
(b) for $4.5\rho_{0}$ and (c) for $7\rho_{0}$.}
\end{figure}

\end{document}